\documentclass{llncs}
\usepackage{amsfonts}
\usepackage{amsmath}
\usepackage{pst-all}

\usepackage{algorithm}
\usepackage{algpseudocode}

\algnewcommand\algorithmicinput{\textbf{Input:}}
\algnewcommand\algorithmicoutput{\textbf{Output:}}
\algnewcommand\Input{\item[\algorithmicinput]}
\algnewcommand\Output{\item[\algorithmicoutput]}

\def\F{\mathbb{F}}

\newgray{lightgray}{0.7}
\newgray{gray1}{0.9}
\newgray{gray2}{0.8}
\newgray{gray3}{0.7}
\newgray{gray4}{0.2}

\begin{document}
%
%
\pagestyle{headings}  
\title{Explicit formulas for efficient \\ multiplication in $\F_{3^{6m}}$}
\titlerunning{Efficient multiplication in $\F_{3^{6m}}$}
\author{Elisa Gorla\inst{1} \and Christoph Puttmann\inst{2} \and Jamshid Shokrollahi\inst{3}}
%
%
%
\institute{
University of Zurich, Switzerland\\
\email{elisa.gorla@math.unizh.ch} \and
Heinz Nixdorf Institute, University of Paderborn, Germany\\
\email{puttmann@hni.upb.de} \and
Ruhr University, Bochum, Germany\\
\email{jamshid@crypto.rub.de}
}

\maketitle              

\begin{abstract}
Efficient computation of the Tate pairing is an important part of
pairing-based cryptography. Recently with the introduction of the Duursma-Lee
method special attention has been given to the fields of characteristic
$3$. Especially multiplication in $\F_{3^{6m}}$, where $m$ is prime, is an
important operation in the above method. In this paper we propose a new method to reduce the number of
$\F_{3^{m}}$-multiplications for multiplication in $\F_{3^{6m}}$ from $18$
in recent implementations to $15$.
The method is based on the fast Fourier transform and its explicit formulas are given. 
The execution times of our software implementations for $\F_{3^{6m}}$ show the
efficiency of our results.
\end{abstract}


\section{Introduction}

Efficient multiplication in finite fields is a central task in the implementation of most public key cryptosystems. A great amount of work has been devoted to this topic (see \cite{knu98} or \cite{gatger03} for a comprehensive list).
The two types of finite fields which are mostly used in cryptographic standards are binary finite fields of type $\F_{2^{m}}$ and prime fields of type
  $\F_{p}$, where $p$ is a prime (cf. \cite{dss00}).
Efforts to efficiently fit finite field arithmetic into commercial processors
  resulted into applications of medium characteristic finite fields like those
  reported in \cite{baipaa98} and \cite{avamih03}.
Medium characteristic finite fields are fields of type $\F_{p^{m}}$, where $p$ is a prime slightly smaller than the word size of the processor, and has a special form that simplifies the modular reduction.
Mersenne prime numbers constitute an example of primes which are used in this context.
The security parameter is given by the length of the binary representations of the
field elements, and the extension degree $m$ is selected appropriately.
Due to security considerations, the extension degree for fields of characteristic $2$ or medium characteristic is usually chosen to be prime.

With the introduction of the method of Duursma and Lee for the computation of the Tate pairing (see \cite{duulee03}), fields of type $\F_{3^{m}}$ for $m$ prime have attracted special attention.
Computing the Tate pairing on elliptic curves defined over $\F_{3^{m}}$ requires computations both in $\F_{3^m}$ and in $\F_{3^{6m}}$.
In the paper \cite{kermar05}, calculations are implemented using the tower of extensions $$\F_{3^m}\subset\F_{3^{2m}}\subset\F_{3^{6m}}.$$
Multiplications in $\F_{3^{2m}}$ and $\F_{3^{6m}}$ are done using $3$ and $6$ multiplications, respectively. This requires a total $18$ multiplications in $\F_{3^{m}}$.
In this paper we make use of the same extension tower, using $3$ multiplications in $\F_{3^{m}}$ to multiply elements in $\F_{3^{2m}}$.
Since we represent the elements of $\F_{{3^{6m}}}$ as polynomials with coefficients in $\F_{3^{2m}}$, we can use Lagrange interpolation to perform the multiplication. This requires only $5$ multiplications in $\F_{3^{2m}}$, thus reducing the total number of $\F_{3^{m}}$ multiplications from $18$ to $15$.
The method that we propose has a slightly increased number of additions in comparison to the Karatsuba method. Notice however that for $m > 90$ (which is the range used in the cryptographic applications) a multiplication in $\F_{3^{m}}$
requires many more resources than an addition, therefore the overall resource consumption is reduced, as also shown by the results of our software experiments shown in Section~\ref{sec:efficient}.

In comparison to the classical multiplication method, the Karatsuba method (see \cite{karofm63}, \cite{paa94}, and \cite{kermar05}) reduces the number of multiplications while introducing extra additions. Since the cost of addition grows linearly in the length of the polynomials, when the degree of the field extension gets larger multiplication will be more expensive than addition. Hence the above tradeoff makes
sense. The negligibility of the cost of addition compared to that of
multiplication has gone so far that the theory of multiplicative
complexity of bilinear maps, especially polynomial multiplication,
takes into account only the number of variable multiplications (see e.g.
\cite{lemwin77} and \cite{win80}). Obviously this theoretical model is
of practical interest only when the number of additions and the
costs of scalar multiplications can be kept small. A famous result in
the theory of multiplicative complexity establishes a lower bound of $2n+1$ for the number of variable multiplications needed for the computation of the product of two
polynomials of degree at most $n$. This lower bound can be achieved only when the field contains enough elements (see \cite{burcla97} or~\cite{bajimb06}).
The proof of the theorem uses Lagrange evaluation-interpolation, which is
also at the core of our approach. This is similar to the short polynomial multiplication (convolution) methods for complex or real numbers in~\cite{bla85}. In order for this method to be especially efficient, the points at which evaluation and interpolations are done are selected as primitive $(2n+1)$st roots of
unity. In a field of type $\F_{3^{2m}}$, fifth roots of unity do not
exist for odd $m$. We overcome this problem by using fourth
roots of unity instead. Notice that a primitive fourth root of unity always exist in a field of type $\F_{3^{2m}}.$ We use an extra point to compute the fifth
coefficient of the product. An advantage of using a primitive fourth root of unity is
that the corresponding interpolation matrix will be a $4\times4$ DFT matrix,
and the evaluations and interpolations can be computed using radix-$2$ FFT
techniques (see \cite{cootuk65} or \cite{loa92}) to save some further number of additions and scalar
multiplications.
The current work can be considered as the continuation of that in \cite{gatsho05} for combination of the linear-time multiplication methods with the classical or Karatsuba ones to achieve efficient polynomial multiplication formulas.

Our work is organized as follows. Section~\ref{sec:lagrange} is devoted to explaining how evaluation-interpolation can be used in general to produce short polynomial multiplication methods. In Section~\ref{sec:rootsofunity} we show how to apply this method to our special case, and produce explicit formulas for multiplication of polynomials of degree at most $2$ over $\F_{3^{2m}}$.
In Section~\ref{sec:efficient} we fine-tune our method using FFT techniques,
and give timing results of software implementations and also explicit multiplication formulas. Section~\ref{sec:anotherexample} shows how our results can be used in conjunction with the method of Duursma-Lee for computing the Tate pairing on some elliptic and hyperelliptic curves.
Section~\ref{sec:conclusion} contains some final remarks and conclusions.

\section{Multiplication using evaluation and interpolation}
\label{sec:lagrange}

We now explain the Lagrange evaluation-interpolation for polynomials
with coefficients in $\F_{p^{m}}$. Throughout this section $m$ is not
assumed to be prime (in the next section we will replace $m$ by $2m$). Let
\begin{align*}
& a(z) =a_{0}+a_{1}z+ \cdots +a_{n}z^{n} \in \F_{p^{m}}[z]\\
& b(z) =a_{0}+a_{1}z+ \cdots +a_{n}z^{n} \in \F_{p^{m}}[z]
\end{align*}
be given such that
\begin{equation}
p^{m} > 2n.
\label{equ:fieldsizecondition}
\end{equation}
We represent the product of the two polynomials by
$$c(z)=a(z)b(z)=c_{0}+c_{1}z+ \cdots +c_{2n}z^{2n}$$
and let $e=(e_{0}, \cdots, e_{2n}) \in \F_{p^{m}}^{2n+1}$ be a
vector with $2n+1$ distinct entries.
Evaluation at these points is given by the map $\phi_{e}$
\begin{align*}
& \phi_{e}: \F_{p^{m}}[z] \rightarrow \F_{p^{m}}^{2n+1}\\
& \phi_{e}(f) = (f(e_{0}), \cdots, f(e_{2n})).
\end{align*}
Let $A, B, C \in \F_{p^{m}}^{2n+1}$ denote the vectors
$(a_{0}, \cdots, a_{{n}}, 0, \cdots, 0)$, \linebreak $(b_{0}, \cdots, b_{{n}}, 0,
\cdots, 0)$, and $(c_{0}, \cdots, c_{2n})$, respectively.
Using the above notation we have
$$\phi_{e}(a)=V_{e}A^{T},\;\;\; \phi_{e}(b)=V_{e}B^{T},\;\;\;\mbox{and}\;\;\; \phi_{e}(c)=V_{e}C^{T},$$
where $V_{e}$ is the Vandermonde matrix
$$V_{e}=\left(
\begin{array}{cccc}
  1 & e_{0} & \cdots & e_{0}^{2n}\\
  1 & e_{1} & \cdots & e_{1}^{2n}\\
  \vdots & \vdots & \ddots & \vdots\\
  1 & e_{2n} & \cdots & e_{2n}^{2n}
\end{array}
\right).$$

The $2n+1$ coefficients of the product $c(z)=a(z) \cdot b(z)$ can be
computed using interpolation applied to the evaluations of $c(z)$ at the chosen
$2n+1$ (distinct) points of $\F_{p^m}$.
These evaluations can be computed by multiplying the
evaluations of $a(z)$ and $b(z)$ at these points.
This can be formally written as $$\phi_e(c)=\phi_{e}(a)*\phi_{e}(b)$$
where we denote componentwise multiplication of vectors by $*$.
Equivalently, if we let $W_e$ be the inverse of the matrix $V_e$, we
have that
$$C^{T}=W_e (\phi_{e}(a)*\phi_{e}(b))$$
which allows us to compute the vector $C$, whose entries are the
coefficients of the polynomial $c(z)$.

When condition (\ref{equ:fieldsizecondition}) is satisfied, the polynomial
multiplication methods constructed in this way have the smallest
multiplicative complexity, i.e. the number of variable multiplications in $\F_{p^m}$ achieves the lower bound $2n+1$ (see~\cite{burcla97}).
Indeed (\ref{equ:fieldsizecondition}) can be relaxed to hold even for
$p^m=2n$.
In this case, a virtual element $\infty$ is added to the finite field.
This corresponds to the fact that the leading coefficient of the product is the product of the leading coefficients of the factors.

Application of this method to practical situations is not straightforward, since
the number of additions increases and eventually dominates the reduction in the number
of multiplications. In order for this method to be efficient, $n$ must be much smaller
than $p^{m}$. An instance of this occurs when computing in extensions of medium size
primes (see e.g.~\cite{bajimb06}). The case of small values of $p$ is more complicated, even for small values of $n$. We recall that in this case the entries of the matrix $V_{e}$ are in $\F_{p^{m}}$ and are generally represented as polynomials of length $m-1$ over $\F_{p}$.
For multiplication of $V_{e}$ by vectors to be efficient, the entries of this
matrix must be chosen to be sparse. However, this gives no control on the sparsity of the entries of $W_{e}$. Indeed one requirement for the entries of $W_{e}$, in the basis $\mathcal{B}$, to be
sparse is that the inverse of the determinant of $V_{e}$, namely
$$\prod_{0 \le i,j \le 2n, i \neq j}(e_{i}-e_{j})$$
has a sparse representation in $\mathcal{B}$. We are not aware of any method which can be used here.
On the other hand, it is known that if the $e_{i}$'s are the elements of the
geometric progression $\omega^{i}$, $0 \le i \le 2n$, and $\omega$ is a
$(2n+1)$st primitive root of unity, then the inverse $W_{e}$ equals $1/(2n+1)$ times the
Vandermonde matrix whose $e_{i}$'s are the elements of the geometric progression of $\omega^{-1}$ (see~\cite{gatger03}). We denote these two matrices by $V_{\omega}$ and
$V_{\omega^{-1}}$, respectively.
The above fact suggests that choosing powers of roots of unity as interpolation points should enable us to control the sparsity of the entries of the corresponding Vandermonde matrix. Roots of unity are used in different contexts for multiplication of polynomials, e.g. in the FFT (see \cite{gatger03}) or for the construction of short multiplication methods in \cite{bla85}.
In the next section we discuss how to use fourth roots of unity to compute multiplication in $\F_{p^{6m}}$, using only $5$ multiplications in $\F_{3^{2m}}$.

\section{Multiplication using roots of unity}
\label{sec:rootsofunity}

Elements of $\F_{3^{6m}}$ can be represented as polynomials of degree at most $2$ over $\F_{3^{2m}}$. Therefore, their product is given by a polynomial of degree at most $4$ with coefficients in $\F_{3^{2m}}$.
In order to use the classical evaluation-interpolation method we would need a primitive fifth root of unity. This would require $3^{2m}-1$ to be a multiple of $5$, and this is never the case unless $m$ is even (recall that cryptographic applications require $m$ to be prime).
However using the relation
\begin{equation}
c_{4}=a_{2}b_{2}
\label{equ:interpolateinfinity}
\end{equation}
we can compute the coefficients of $c(x)$ via
\begin{equation}\label{equ:2nfirst}
\left(
\begin{array}{cccc}
  1 & 1 & 1 & 1\\
  1 & \omega & \omega^{2} & \omega^{3}\\
  1 & \omega^{2} & 1 & \omega^{2} \\
  1 & \omega^{3} & \omega^{2} & \omega
\end{array}
\right)\left(\begin{array}{c}c_0 \\ c_1 \\ c_{2} \\ c_{3}
  \end{array}\right)
=\left(\begin{array}{c}
a(1)b(1)-c_{4} \\
a(\omega)b(\omega)-c_{4} \\
a(\omega^{2})b(\omega^{2})-c_{4} \\
a(\omega^{3})b(\omega^{3})-c_{4}
\end{array}\right)\end{equation}
where $\omega$ is a fourth root of unity.
Now we apply (\ref{equ:interpolateinfinity}) and (\ref{equ:2nfirst}) to find
explicit formulas for multiplying two polynomials of degree at most $2$ over
$\F_{3^{2m}}$, where $m>2$ is a prime.

We follow the tower representation of \cite{kermar05}, i.e.
\begin{equation}
	\begin{array}{l}
 \F_{3^{m}} \cong \F_{{3}}[x]/(f(x))\\
 \F_{3^{2m}} \cong \F_{3^{m}}[y]/(y^2+1)\\
	\end{array}
\label{equ:towerF2}
\end{equation}
where $f(x)\in\F_3[x]$ is an irreducible polynomial of degree $m$.
Denote by $s$ the equivalence class of $y$.
Note that for odd $m > 2$, $4 \not | 3^m-1$ and hence $y^2+1$ is irreducible over $\F_{3^m}$ since the roots of $y^2+1$ are fourth roots of unity.
Let
\begin{equation}
a(z)=a_{0}+a_{1}z+a_{2}z^{2},\;\;\;\; b(z)=b_{0}+b_{1}z+b_{2}z^{2}
\label{equ:f36mrep}
\end{equation}
be polynomials in $\F_{p^{3^{2m}}}[z]^{\le 2}.$
Our goal is computing the coefficients of the polynomial
$$c(z)=a(z)b(z)=c_{0}+c_{1}z+\cdots c_{4}z^{4}.$$
Evaluation of $a(z)$ and $b(z)$ at
$(1,s,s^{2},s^{3})=(1,s,-1,-s)$ can be done by multiplying the Vandermonde matrix of
powers of $s$
\begin{equation}
V_{s}=\left(
\begin{array}{cccc}
1 & 1 & 1 & 1\\
1 & s & -1 & -s\\
1 & -1 & 1 & -1\\
1 & -s & -1 & s
\end{array}
\right)
\label{equ:Vs}
\end{equation}
by the vectors $(a_{0}, a_{1}, a_{2}, 0)^{T}$ and $(b_{0}, b_{1}, b_{2},
0)^{T}$, respectively. This yields the vectors
$$\phi_{e}(a)=\left(
\begin{array}{c}
a_{0}+a_{1}+a_{2}\\
a_{0}+sa_{1}-a_{2}\\
a_{0}-a_{1}+a_{2}\\
a_{0}-sa_{1}-a_{2}
\end{array}
\right)\;\;\;\;\mbox{and}\;\;\;\;
\phi_{e}(b)=\left(
\begin{array}{c}
b_{0}+b_{1}+b_{2}\\
b_{0}+sb_{1}-b_{2}\\
b_{0}-b_{1}+b_{2}\\
b_{0}-sb_{1}-b_{2}
\end{array}
\right).$$
Let $\phi_{e}(c)=\phi_e(a)*\phi_e(b)$ be the componentwise product of $\phi_e(a)$ and $\phi_e(b)$
$$
\phi_{e}(c)=
\left(
\begin{array}{c}
P_{0}\\
P_{1}\\
P_{2}\\
P_{3}
\end{array}
\right)=
\left(
\begin{array}{c}
(a_{0}+a_{1}+a_{2})(b_{0}+b_{1}+b_{2})\\
(a_{0}+sa_{1}-a_{2})(b_{0}+sb_{1}-b_{2})\\
(a_{0}-a_{1}+a_{2})(b_{0}-b_{1}+b_{2})\\
(a_{0}-sa_{1}-a_{2})(b_{0}-sb_{1}-b_{2})
\end{array}
\right).
$$
Using (\ref{equ:interpolateinfinity}) and (\ref{equ:2nfirst}) we get
$$
\left(
\begin{array}{c}
c_{0}\\
c_{1}\\
c_{2}\\
c_{3}
\end{array}
\right)=W_{s}
\left(
\begin{array}{c}
P_{0}-P_{4}\\
P_{1}-P_{4}\\
P_{2}-P_{4}\\
P_{3}-P_{4}
\end{array}
\right),
$$
where $P_{4}=a_{2}b_{2}$ and
\begin{equation}
W_{s}=V_{s}^{-1}=\left(
  \begin{array}{cccc}
1 & 1 & 1 & 1 \\
1 & -s & -1 & s \\
1 & -1 & 1 & -1 \\
1 & s & -1 & -s
  \end{array}
\right)
\label{equ:Ws}
\end{equation}
Thus the explicit formulas for the coefficients of the product are
\begin{equation}
\begin{array}{l}
c_{0}=P_{0}+P_{1}+P_{2}+P_{3}-P_{4}\\
c_{1}=P_{0}-sP_{1}-P_{2}+sP_{3}\\
c_{2}=P_{0}-P_{1}+P_{2}-P_{3}\\
c_{3}=P_{0}+sP_{1}-P_{2}-sP_{3}\\
c_{4}=P_{4}.
\end{array}
\label{equ:explicitformulas}
\end{equation}

\section{Efficient implementation}
\label{sec:efficient}

We owe the efficiency of our method to the Cooley-Tukey factorization of the
DFT matrix (\cite{cootuk65}). The matrices $V_{s}$ and $W_{s}$ in
(\ref{equ:Vs}) and (\ref{equ:Ws}) are not sparse, but they are the DFT matrices
of the fourth roots of unity $s$ and $s^{3}$, respectively. Hence they can be
factored as a product of two sparse matrices as shown in (\ref{equ:VsFactor}) and (\ref{equ:WsFactor}).
\begin{align}
& \label{equ:VsFactor}V_{s} = \left(
  \begin{array}{cccc}
    1 & 1 & 1 & 1\\
    1 & s & -1 & -s\\
    1 & -1 & 1 & -1\\
    1 & -s & -1 & s
  \end{array}
\right)=
&\left(
  \begin{array}{cccc}
    1 & 1 & 0 & 0\\
    0 & 0 & 1 & s\\
    1 & -1 & 0 & 0\\
    0 & 0 & 1 & -s
  \end{array}
\right)
\left(
  \begin{array}{cccc}
    1 & 0 & 1 & 0\\
    0 & 1 & 0 & 1\\
    1 & 0 & -1 & 0\\
    0 & 1 & 0 & -1
  \end{array}
\right),\\
& \label{equ:WsFactor}W_{s}=
\left(
  \begin{array}{cccc}
1 & 1 & 1 & 1 \\
1 & -s & -1 & s \\
1 & -1 & 1 & -1 \\
1 & s & -1 & -s
  \end{array}
\right)=
&\left(
  \begin{array}{cccc}
    1 & 1 & 0 & 0\\
    0 & 0 & 1 & -s\\
    1 & -1 & 0 & 0\\
    0 & 0 & 1 & s\\
  \end{array}
\right)
\left(
  \begin{array}{cccc}
    1 & 0 & 1 & 0\\
    0 & 1 & 0 & 1\\
    1 & 0 & -1 & 0\\
    0 & 1 & 0 & -1
  \end{array}
\right).
\end{align}
The factorizations in (\ref{equ:VsFactor}) and (\ref{equ:WsFactor})
allow us to efficiently compute the product of the matrices $V_{s}$
and $W_{s}$ with vectors.
Notice also that the product of an element
$\omega = us+v \in \F_{3^m}[s]^{\leq 1}\cong\F_{3^{2m}}$ with $s$
equals $vs-u$. Hence multiplying by $s$ an element of $\F_{3^{2m}}$ is
not more expensive than a change of sign.

Notice that in alternative to the Vandermonde matrix corresponding to
$s$ we could use the matrix
$$\left(
\begin{array}{cccc}
1 & 0 & 0 & 0 \\
1 & 1 & 1 & 1 \\
1 & -1 & 1 & -1 \\
1 & s & -1 & -s
\end{array}
\right)$$
whose inverse is
$$\left(
\begin{array}{cccc}
1 & 0 & 0 & 0 \\
s & 1-s & -1-s & s \\
-1 & -1 & -1 & 0 \\
-s & 1+s & -1+s & -s
\end{array}
\right).$$
Obviously the latter matrices are sparse but since they do
not possess any special structure up to our knowledge, multiplying
them by vectors is more expensive than multiplying $V_{s}$ and
$W_{s}$.

Multiplying elements in the field $\F_{3^{6\cdot 97}}$ is
required in the Tate pairing computation on the group of
$\F_{3^{97}}$-rational points of the elliptic curves
$$E_d: y^2=x^3-x+d\;\;\; d\in\{-1,1\}$$ defined over $\F_3$. An efficient
algorithm for the computation of the Tate pairings on these curves
is discussed in \cite{duulee03}.

We have implemented the multiplication over $\F_{3^{6 \cdot 97}}$ using the Karatsuba method, the Montgomery method from
\cite{mon05}, and our proposed method on a PC with an AMD Athlon 64
processor 3500+. 
The processor was running at 2.20~GHz and we have used the NTL library (see \cite{ntl06})
for multiplication in $\F_{3^{97}}$. 
Please note that although we have chosen $m=97$ for benchmarking purposes, these methods can be applied to any odd $m > 2$ as mentioned in Section~\ref{sec:rootsofunity}.

\begin{table}[htbp]
 \centering
 \begin{tabular}{|l|c|} \hline
   {\bf Multiplication method} & {\bf Elapsed time (ms)}\\ \hline
   Karatsuba method & $1.698$\\
   Montgomery method & $1.605$\\
   Proposed  method & $1.451$\\ \hline
 \end{tabular}
 \caption{Comparison of the execution times of the Karatsuba and Montgomery
   multipliers with the proposed method for $\F_{3^{6m}}$.}
 \label{tab:runningtime}
\end{table} 

The execution times are shown in Table~\ref{tab:runningtime}. 
For the
Karatsuba and the proposed methods we have used the tower of extensions
$$
\F_{3^{97}} \subset \F_{3^{2 \cdot 97}} \subset \F_{3^{6 \cdot 97}},
$$
where
\begin{align*}
\F_{3^{97}} & \cong \F_{{3}}[x]/(x^{97}+x^{16}+2)\\
\F_{3^{2 \cdot 97}} & \cong \F_{3^{97}}[y]/(y^2+1)\\
\F_{3^{6 \cdot 97}} & \cong \F_{3^{2 \cdot 97}}[z]/(z^{3}-z-1),
\end{align*}
whereas for the Montgomery method the representation
$$
\F_{3^{6 \cdot 97}} \cong \F_{3^{97}}[y]/(y^{6}+y-1)
$$
has been used. Our implementations show that the new method is almost $14$\% faster than
the Karatsuba and $10$\% faster than the Montgomery method, which is almost the
ratio of saved multiplications. This provides further evidence for the
fact that the number
of multiplications in $\F_{3^{97}}$ is a good indicator of the performance of the method
for $\F_{3^{6 \cdot 97}}$.

Our multiplications are based on the following formulas.
Let $\alpha, \beta \in \F_{3^{6 \cdot m}}$ be given as:
\begin{align*}
& \alpha =a_0 + a_1s + a_2r + a_3rs + a_4r^2 + a_5r^2s,\\
& \beta = b_0 + b_1s + b_2r + b_3rs + b_4r^2 + b_5r^2s,
\end{align*}
where $a_0, \cdots , b_5 \in \F_{3^m}$ and $s \in F_3^{2 \cdot m}$ , $r \in \F_3^{6 \cdot m}$ are roots of
$y^2 + 1$ and $z^3-z-1$, respectively. Let their product 
$\gamma = \alpha \beta \in \F_{3^{6 \cdot m}}$  be
$$
\gamma = c_0 + c_1s + c_2r + c_3rs + c_4r^2 + c_5r^2s.
$$
The coefficients $c_i$, for $0 \le i \le 5$ are computed using:
$$
\begin{array}{l}
P_0 = (a_0 + a_2 + a_4)(b_0 + b_2 + b_4)\\
P_1 = (a_0 + a_1 + a_2 + a_3 + a_4 + a_5)(b_0 + b_1 + b_2 + b_3 + b_4 + b_5)\\
P_2 = (a_1 + a_3 + a_5)(b_1 + b_3 + b_5)\\
P_3 = (a_0 - a_3 - a_4)(b_0 - b_3 - b_4)\\
P_4 = (a_0 + a_1 + a_2 - a_3 - a_4 - a_5)(b_0 + b_1 + b_2 - b_3 - b_4 - b_5)\\
P_5 = (a_1 + a_2 - a_5)(b_1 + b_2 - b_5)\\
P_6 = (a_0 - a_2 + a_4)(b_0 - b_2 + b_4)\\
P_7 = (a_0 + a_1 - a_2 - a_3 + a_4 + a_5)(b_0 + b_1 - b_2 - b_3 + b_4 + b_5)\\
P_8 = (a_1 - a_3 + a_5)(b_1 - b_3 + b_5)\\
P_9 = (a_0 - a_3 - a_4)(b_0 + b_3 - b_4)\\
P_{10} = (a_0 + a_1 - a_2 + a_3 - a_4 - a_5)(b_0 + b_1 - b_2 + b_3 - b_4 - b_5)\\
P_{11} = (a_1 - a_2 - a_5)(b_1 - b_2 - b_5)\\
P_{12} = a_4b_4\\
P_{13} = (a_4 + a_5)(b_4 + b_5)\\
P_{14} = a_5b_5\\
c_0 = -P_0 + P_2 - P_3  - P_4  + P_{10} + P_{11} - P_{12} + P_{14};\\
c_1 = P_0 - P_1 + P_2 + P_4 + P_5 + P_9 + P_{10} + P_{12} - P_{13} + P_{14}\\
c_2 = -P_0 + P_2 + P_6 - P_8 + P_{12} - P_{14}\\
c_3 = P_0 - P_1 + P_2 - P_6 + P_7 - P_8 - P_{12} + P_{13} - P_{14}\\
c_4 = P_0 - P_2 - P_3  + P_5 + P_6 - P_8 - P_9+ P_{11} + P_{12} - P_{14}\\
c_5 = - P_0 + P_1  - P_2 + P_3 - P_4 + P_5 - P_6 + P_7 - P_8 + P_9- P_{10} + \\P_{11} - P_{12} + P_{13} - P_{14}\\
\end{array}
$$

\section{Other applications of the proposed method}
\label{sec:anotherexample}

Consider the family of hyperelliptic curves
\begin{equation}\label{Cd}
C_d:\; y^2=x^p-x+d\;\;\; d\in\{-1,1\}
\end{equation}
defined over $\F_p$, for $p=3$ mod. $4$. Let $m$ be such that
$(2p,m)=1$ (in practice $m$ will often be prime), and consider the
$\F_{p^m}$-rational points of the Jacobian of $C_d$. An efficient
implementation of the Tate pairing on these groups is given by Duursma
and Lee in \cite{duulee03} and \cite{duulee05}, where they extend analogous results of
Barreto et. al. and of Galbraith et. al. for the case $p=3$. Notice
that this
family of curves includes the elliptic curves $E_d$ that we mentioned in the
last section.
In the aforementioned papers it is
also shown that the curve $C_d$ has embedding degree $2p$. In order to
compute the Tate pairing on this curve, one works with the tower of
field extensions
$$\F_{p^m}\subset\F_{p^{2m}}\subset\F_{p^{2pm}}$$
where the fields are represented as
$$\F_{p^{2m}}\cong\F_{p^m}[y]/(y^2+1)\;\;\;\mbox{and}\;\;\;
\F_{p^{2pm}}\cong\F_{p^{2m}}[z]/(z^p-z+2d).$$
Let $a(z),b(z)\in\F_{p^{2pm}}[z]^{\leq p-1}$, $$a(z)=a_0+a_1z+\ldots+a_{p-1}z^{p-1},$$
$$b(z)=b_0+b_1z+\ldots+b_{p-1}z^{p-1}.$$ Then $c(z)=a(z)b(z)$ has $2p-1$
coefficients, two of which can be computed as $$c_0=a_0b_0\;\;\;
\mbox{and}\;\;\;
c_{2(p-1)}=a_{2(p-1)}b_{2(p-1)}.$$ In order to determine the remaining
$2p-3$ coefficients, we can write a Vandermonde matrix with entries in
$\F_{p^{2m}}^*$ using, e.g., the elements
$$1,2,\ldots,p-1,\pm s,\ldots,\pm\frac{p-3}{2}s,\frac{p-1}{2}s.$$

Another option is writing a Vandermonde matrix using a primitive \linebreak
$2(p-1)$-st root of unity combined with the relation:
$$
c_{2(p-1)}=a_{2(p-1)}b_{2(p-1)}.
$$
Notice that there is an element of order
$2(p-1)$ in $\F_{p^2}$, since $2(p-1)|p^2-1$. If $a$ is a primitive
element in $\F_{p^2}$, then $\omega=a^{(p+1)/2}$ is a primitive $2(p-1)$st
root of unity.

\section{Conclusion}
\label{sec:conclusion}

In this paper we derived new formulas for multiplication in
$\F_{3^{6m}}$, which use only $15$ multiplications in
$\F_{3^{m}}$. Being able to efficiently multiply elements in
$\F_{3^{6m}}$ is a central task for the computation of the Tate
pairing on elliptic and hyperelliptic curves.
Our method is based on the fast Fourier transform, slightly modified to
be adapted to the finite fields that we work on.
Our software experiments show that this method is at least $10$\% faster than
other proposed methods in the literature. We have also discussed use
of these ideas in conjunction with the general methods of Duursma-Lee for
Tate pairing computations on elliptic and hyperelliptic curves.

\section*{Acknowledgement}

The research described in this paper was funded in part by the Swiss National Science Foundation, registered there under grant number 107887, and
by the German Research Foundation (Deutsche Forschungsgemeinschaft
 DFG) under project RU 477/8.
We thank also the reviewers for their precise comments.

\setlength{\unitlength}{1cm}

\end{document}